\documentclass[12pt,preprint,one-column]{emulateapj}

\usepackage{graphicx}
\usepackage{amsmath}
\usepackage{color}
\usepackage{multirow}

\slugcomment{for ApJ}

\shortauthors{Takiwaki, Kotake, and Suwa}

\begin{document}

\title{A Comparison of Two- and Three-dimensional Neutrino-hydrodynamics
 simulations of Core-collapse Supernovae}

\author{Tomoya Takiwaki\altaffilmark{1}, Kei Kotake\altaffilmark{2,1}, 
and Yudai Suwa\altaffilmark{3}}
\affil{\altaffilmark{1}Center for Computational Astrophysics, National
Astronomical Observatory of Japan, 2-21-1, Osawa, Mitaka, Tokyo,
181-8588, Japan}
\affil{\altaffilmark{2}Department of Applied Physics, Fukuoka University, Fukuoka 814-0180, Japan}
\affil{$^3$Yukawa Institute for Theoretical Physics, Kyoto
  University, Oiwake-cho, Kitashirakawa, Sakyo-ku, Kyoto, 606-8502,
  Japan}

\begin{abstract}
We present numerical results on two- (2D) and three-dimensional (3D) hydrodynamic
core-collapse simulations of an 11.2$M_\odot$ star.
By changing numerical resolutions and seed perturbations systematically,
we study how the postbounce dynamics is different in 2D and 3D. The calculations were performed with
 an energy-dependent treatment of the neutrino
 transport based on the isotropic diffusion source
 approximation scheme, which we have updated to achieve a very high computational efficiency.
 All the computed models in this work
 including nine 3D models and fifteen 2D models exhibit the
 revival of the stalled bounce shock, leading to the possibility of
 explosion.
  All of them are driven by the neutrino-heating mechanism, which is
 fostered by neutrino-driven convection and the
 standing-accretion-shock instability (SASI). Reflecting the stochastic
 nature of multi-dimensional (multi-D) neutrino-driven explosions, the blast morphology changes from models to models.
 However, we find that the final fate of the multi-D models whether an explosion is obtained or not, is little affected
      by the explosion stochasticity. In agreement with some 
 previous studies, higher numerical
     resolutions lead to slower onset of the shock
     revival in both 3D and 2D.
    Based on the self-consistent supernova models leading to the possibility of
 explosions, our results systematically show that the revived shock expands more 
energetically in 2D than in 3D.
\end{abstract}


\section{Introduction}
Ever since the first numerical simulation of core-collapse 
 supernovae (CCSNe) \citep{colgate}, 
the neutrino-driven mechanism
 has been the leading candidate of the explosion mechanism 
 for more than four decades. In the long history, a very
 important lesson we have learned from
 \citet{rampp00,lieb01,thom03,sumi05} near the Millennium, 
is that the spherically-symmetric (1D) form of this mechanism 
fails to explode canonical massive stars.
Supported by accumulating supernova observations of the blast morphology
 (e.g., \citet{wang08}, and references therein), 
a number of multi-dimensional (multi-D)
hydrodynamic simulations have been reported so far, which gives 
 us a confidence that hydrodynamic motions associated with 
neutrino-driven convection 
(e.g., \citet{herant,burr95,jankamueller96,frye02} and see
 collective references in \citet{burrows12,murphy13,couch13}) and the 
 SASI (e.g., \citet{blon03,Scheck04,scheck06,Ohnishi06,ohnishi07,ott_multi,murphy08,thierry,thierry07,thierry12,iwakami1,iwakami2,iwakami13,endeve10,endeve12,rodrigo09_2,rodrigo09,rodrigo10,hanke12}) can help the onset of 
neutrino-driven explosions.

In fact, a growing number of neutrino-driven explosions have been 
recently obtained in the state-of-the-art two-dimensional (2D) 
simulations, in which spectral neutrino transport is solved with different
  levels of sophistication (e.g., \citet{buras06,marek,bernhard12,bernhard13,bruenn13,suwa10,suwa11,suwa13}, \citet{janka12} for a review). 
This success is, however, accompanying new questions.
Among them\footnote{Note that general 
relativity (GR, e.g., \citet{bernhard12,kuroda12,ott13}) and detailed weak interactions (e.g., \citet{langanke08,furusawa13}) are 
 considered as important as 3D effects.}, 
three-dimensional (3D) effects on the neutrino-driven mechanism are 
 attracting a paramount attention (e.g., \citet{burrows13,
kotake12} for a review). Unfortunately, however, experimental 
3D models that employed a light-bulb scheme 
(e.g., \citet{murphy08}), 
have provided divergent results so far. 
The basic result of \citet{nordhaus} who were the first 
to point out that 3D leads to easier explosions than 2D, has been 
supported by the follow-up studies \citep{burrows12,dolence12}, 
but not by \citet{hanke12,couch13}. On top of the urgent task to make
a detailed comparison between these idealized models, 
 self-consistent 3D simulations should be done in order to have
 the final word on the 3D effects.

At present, 3D CCSN simulations including spectral neutrino transport 
are only few \citep{hanke13,takiwaki12}. Very recently,
 \citet{hanke13} succeeded in performing 3D simulations with
  detailed neutrino transport for a 
 27 $M_{\sun}$ star. In addition to the first discovery regarding the violent SASI 
activity in self-consistent 3D models, their 
 results illuminate the importance to go beyond 
the prevalent light-bulb scheme, only by doing so, the non-linear
 couplings such as between core-contraction of the proto-neutron star 
(PNS), the accretion neutrino luminosity, and 
the multi-D hydrodynamic feedback of neutrino-driven convection 
and the SASI, can be self-consistently determined. On the other hand,
the very high computational cost allowed \citet{hanke13} 
to focus on a single (self-consistent) 3D model, and it has not been clarified yet whether 
3D helps or harms the onset of neutrino-driven explosions compared to 2D. 

To address this question, we investigate in this paper how the explosion
 dynamics will differ from 3D to 2D by systematically 
  changing numerical resolutions and initial seed perturbations 
  in multi-D radiation-hydrodynamic simulations.
For the multi-group neutrino transport, the isotropic diffusion source
 approximation (IDSA) scheme \citep{idsa} is
 implemented in a ray-by-ray manner (e.g., \citet{takiwaki12} for more
  details), which we have updated to achieve a very high computational 
efficiency.
As in \citet{takiwaki12}, we here focus on the evolution of an 
$11.2 M_{\odot}$ star of \citet{woos02}. We choose this lighter 
 progenitor because the shock revival occurs relatively earlier after 
bounce (e.g., \citet{buras06}) compared 
to more massive progenitor models as employed in \citet{hanke13}.
 The updated transport scheme together with the employed 
earlier-to-explode progenitor allow us to conduct a systematic numerical 
study, for the first time, in both 2D and 3D 
(to the best of our knowledge) in the context of 
self-consistent neutrino-driven supernova models.
 
\begin{figure}[tbh]
\centering
\includegraphics[width=.9\linewidth]{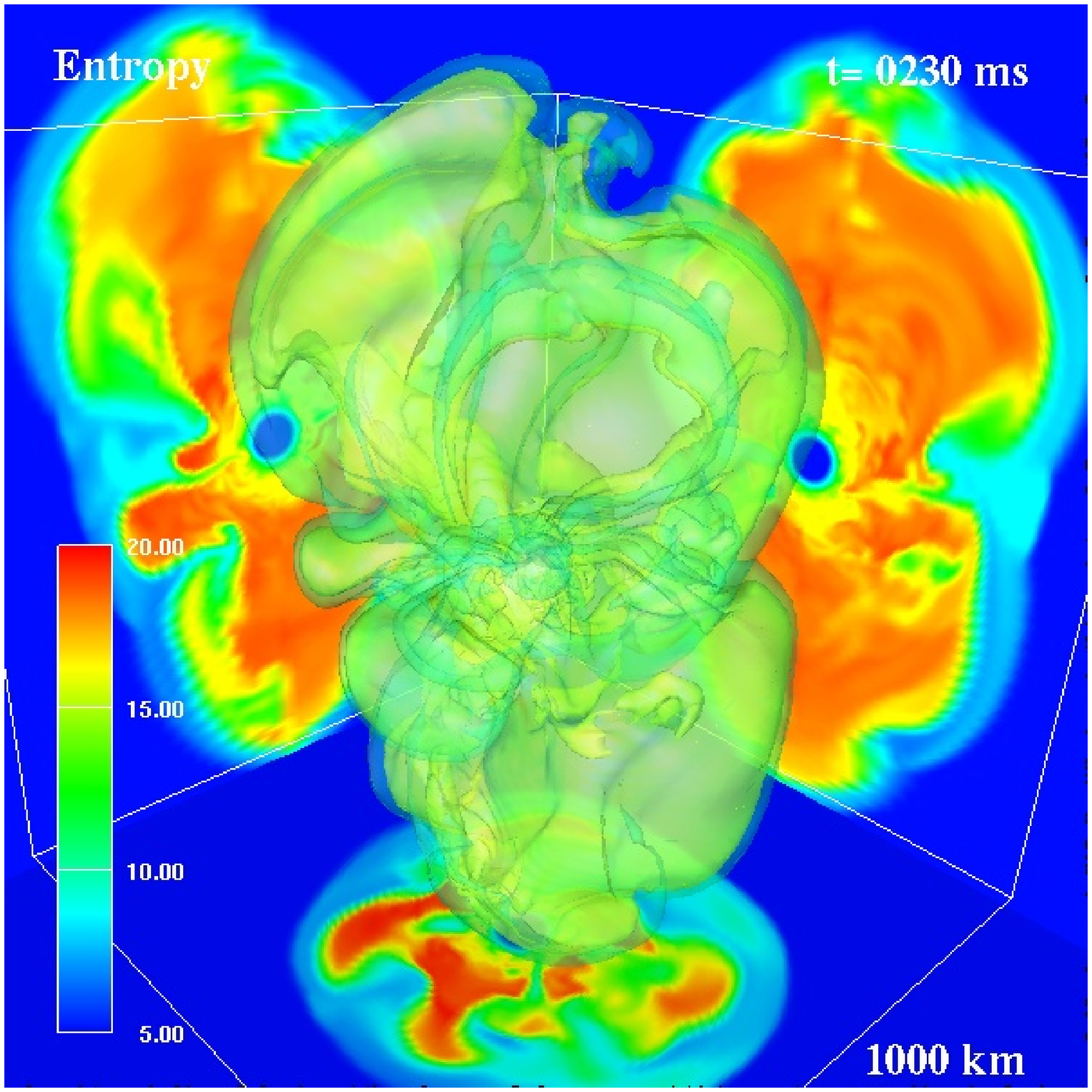}\\
\includegraphics[width=.45\linewidth]{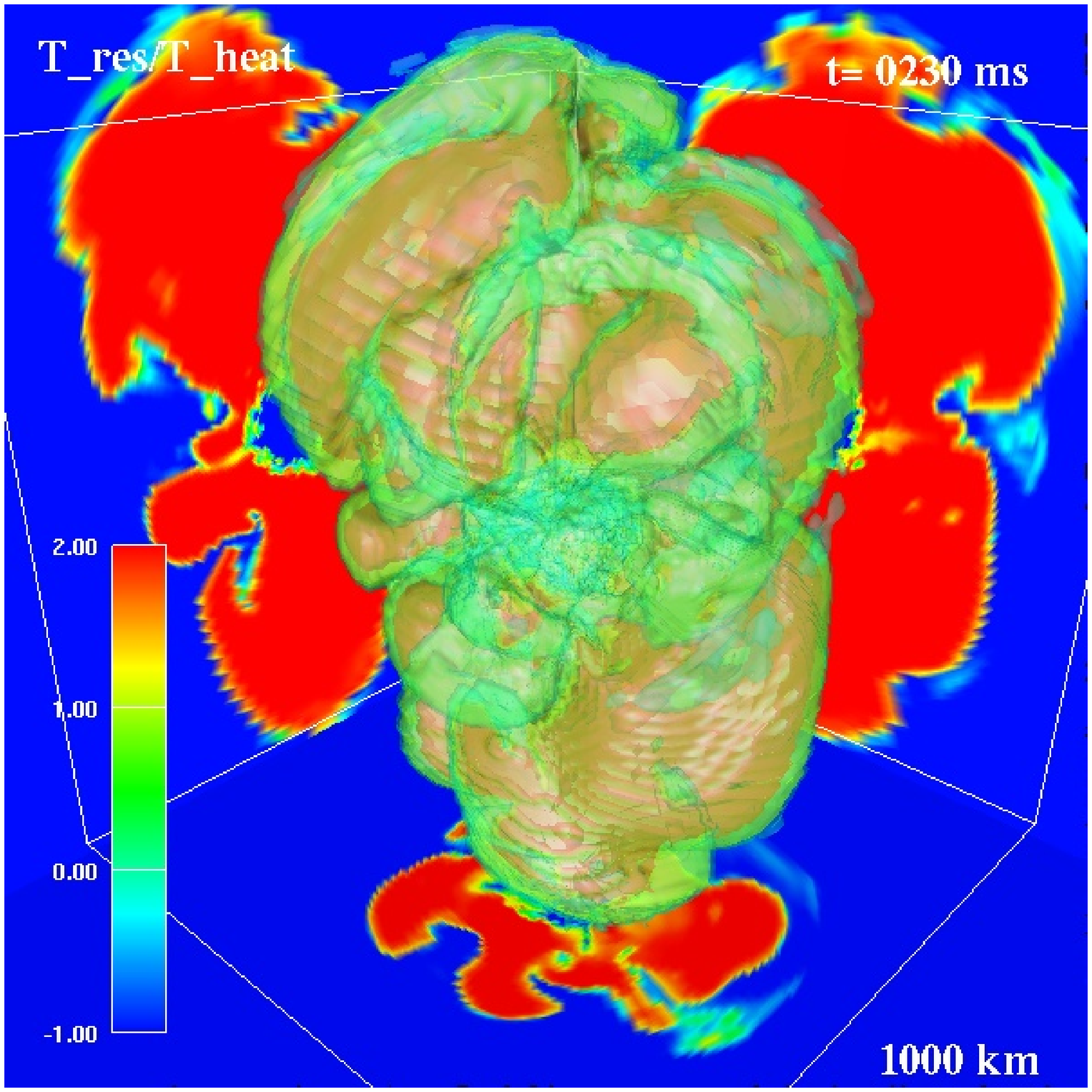}
\includegraphics[width=.45\linewidth]{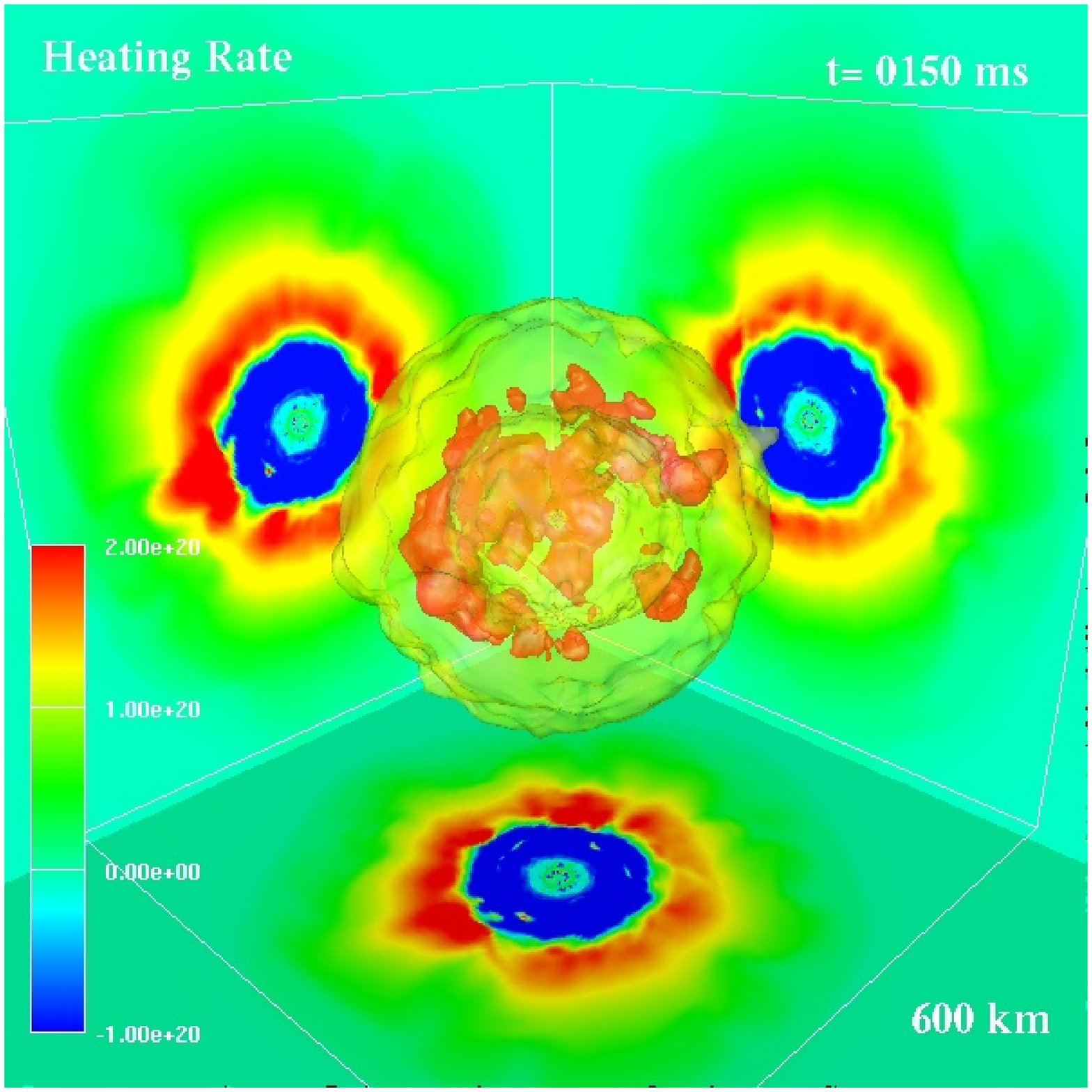}
\caption{Three dimensional plots of entropy per baryon (top panel),
 $\tau_{\rm res}/\tau_{\rm heat}$ (bottom left panel) that is the
 ratio of the residency to the neutrino heating timescale (see
  the text for details), and
  the net neutrino heating rate (bottom right panel, in unit
 of ${\rm erg}~{\rm cm}^{-3}~{\rm s}^{-1}$) for three snapshots
 (top and bottom left: $t=230$ ms, and bottom right: $t=150$ ms measured
  after bounce ($t\equiv 0$) of our model 3D-H-1). 
The contours on the
 cross sections in the $x = 0$ (back right), $y = 0$ (back bottom),
  and $z = 0$ (back left) planes are, respectively, projected on
  the sidewalls of the graphs.
 For each snapshot,
 the length of white line is indicated at right bottom text.} 
\label{fig1}
\end{figure}

\begin{figure*}[htb]
\centering
\includegraphics[width=.28\linewidth]{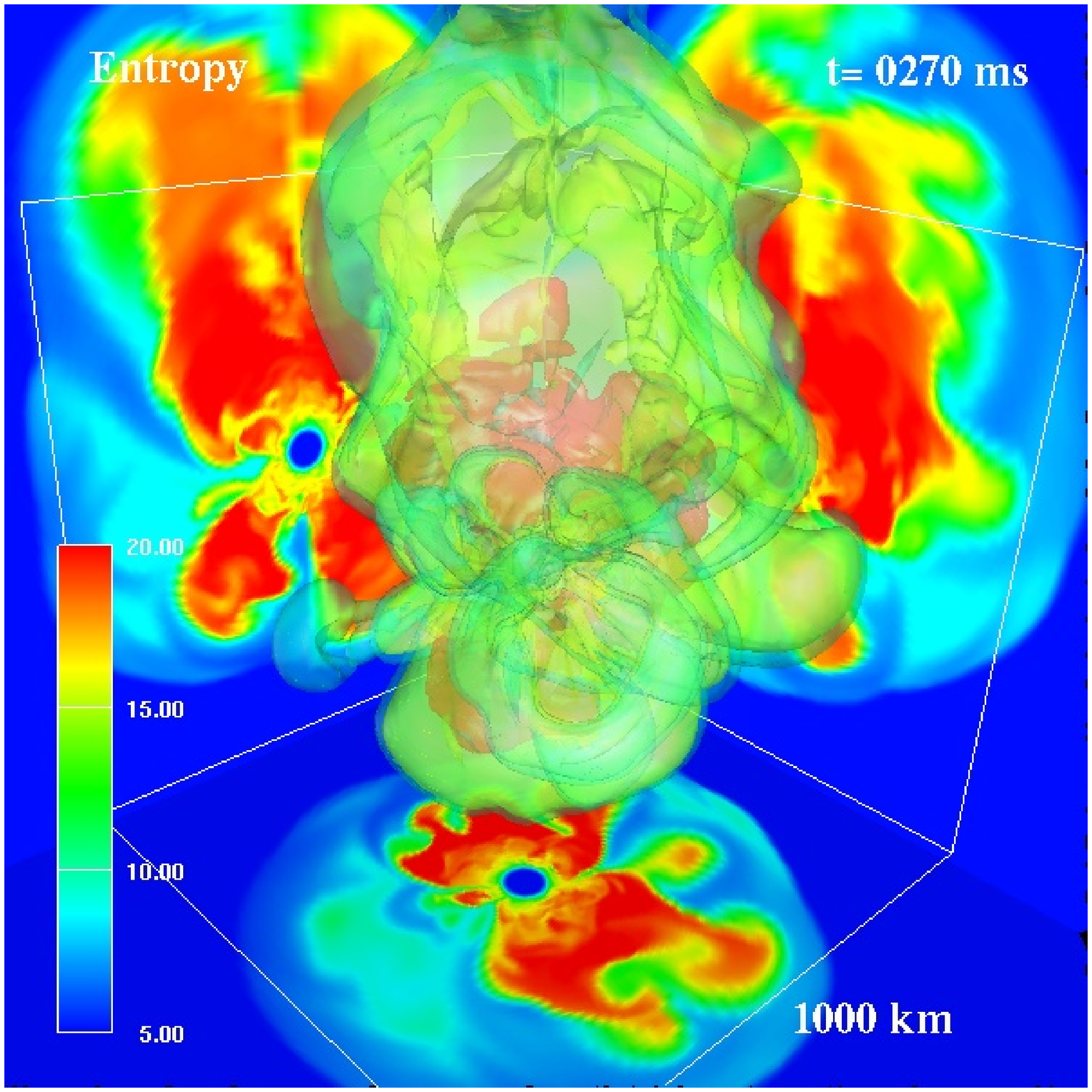}
\includegraphics[width=.28\linewidth]{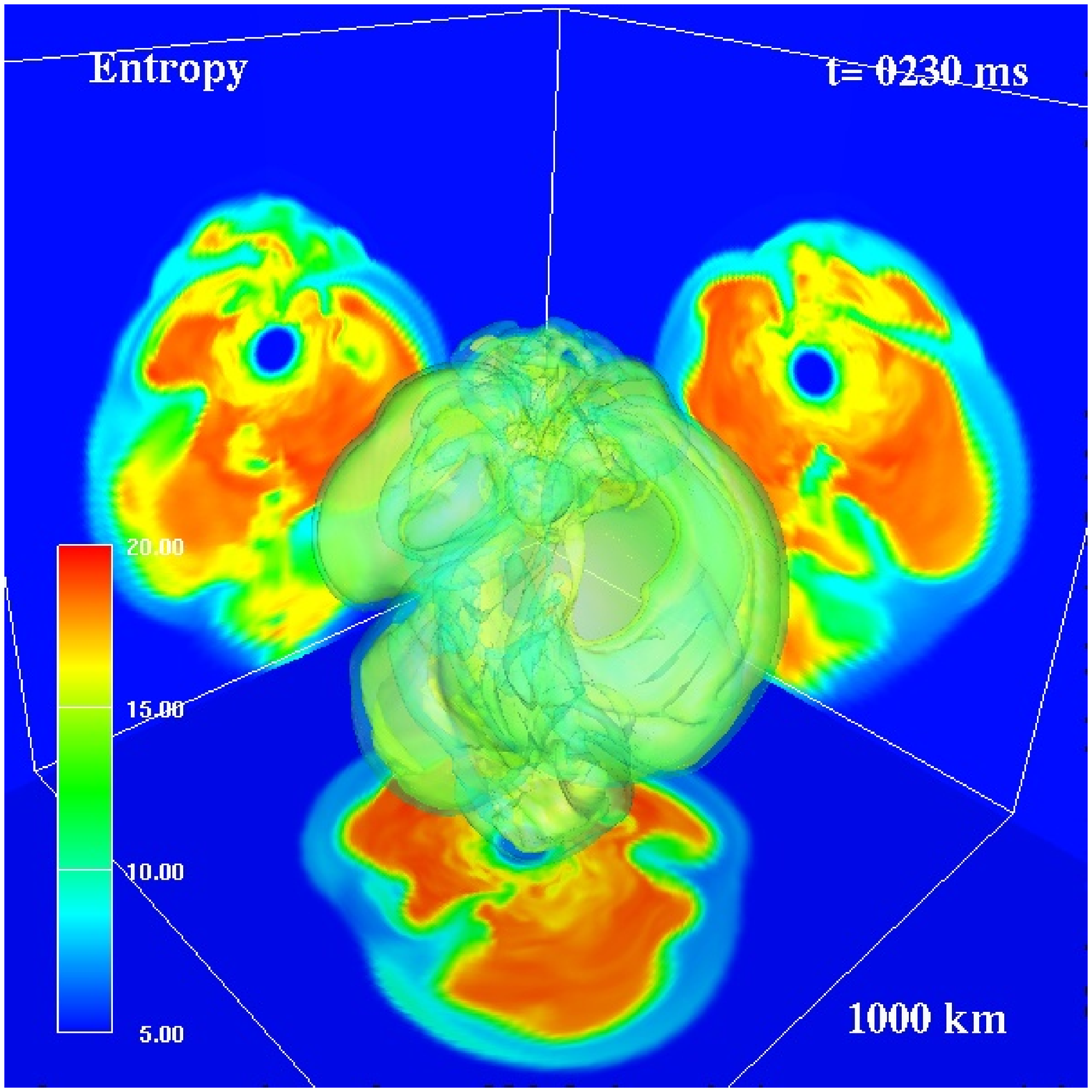}
\includegraphics[width=.4\linewidth]{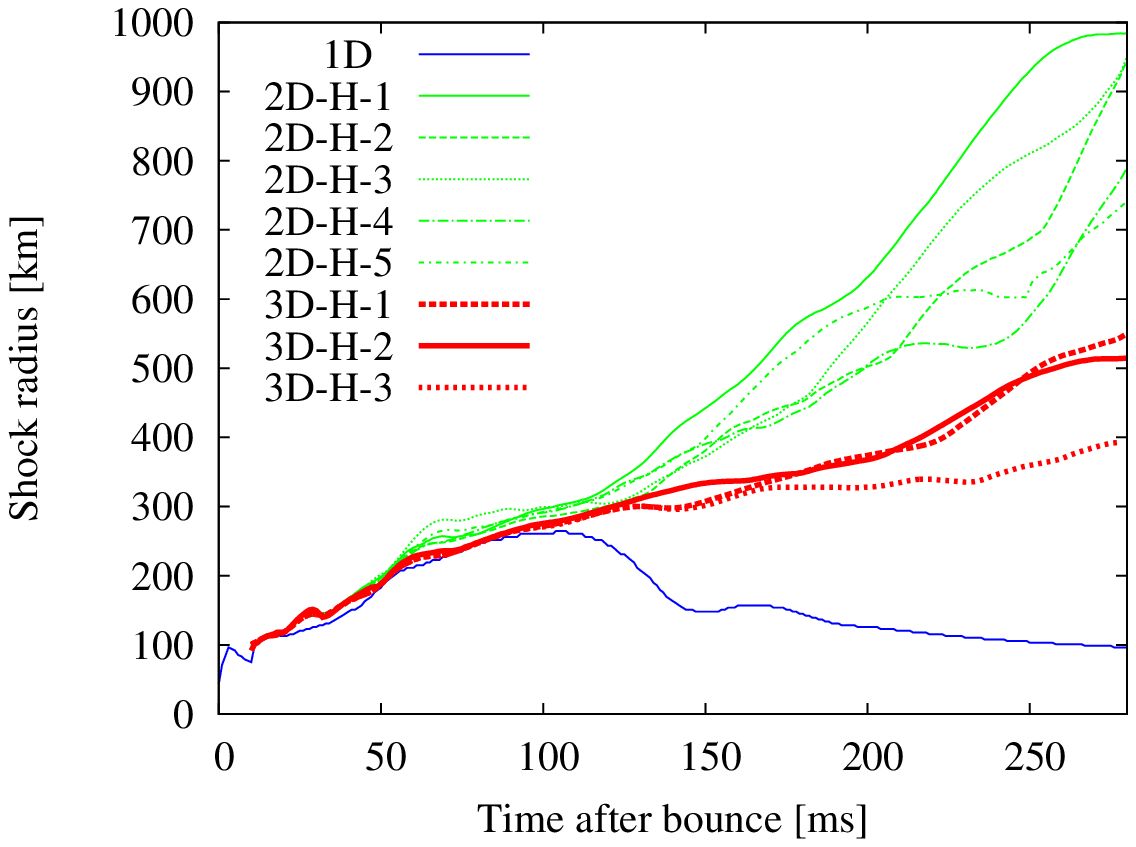}
 \caption{Same as the top panel in Figure \ref{fig1} but for models
 3D-H-2 (left panel) and 3D-H-2 (middle panel), which produces
 stronger explosions closely toward
  the north (left panel) and south pole (middle panel), respectively.
 Right panel shows the evolution of average shock radii for the
 high-resolution 2D
 (green lines) and 3D (red lines) models explored in this study (e.g., Table 1).}\label{fig2}
\end{figure*}

\section{Numerical Methods and Models}

Here we briefly summarize several major updates of the code that
 we have implemented after our previous work \citep{takiwaki12}
 in which the spectral neutrino transport scheme IDSA \citep{idsa} was
implemented in the ZEUS-MP code \citep{hayes}. 

In the original IDSA scheme, a steady-state approximation 
($\partial {f^{\rm s}}/(\partial t) = 0$) is assumed. Here
 $f^{\rm s}$ represents the {\it streaming} part of the neutrino 
distribution function (e.g., \citet{idsa}).
Then one should deal with
a Poisson-type equation to find the solution of $f^s$ 
(e.g., Eq.(10) in \citet{idsa}). This is computationally expensive,
 because a collective data-communication is required
 on the MPI routines for all the processors (along the given radial direction in the ray-by-ray approximation) to solve the Cauchy problem.

To get around the problem,
 one needs to solve the evolution of $f^{\rm s}$ as,
\begin{eqnarray}
\frac{\partial \mathcal{E}^{\rm s}}{c\partial {t}} &+&
\frac{1}{r^2}\frac{\partial}{\partial r}r^2 \mathcal{F}^{\rm s}
= \mathcal{S}[{j}, {\chi}, {\Sigma}],\label{eq:fs-evol}\\
\mathcal{E}^{\rm s} &\equiv& \frac{1}{2}\int d{\mu}\,{f}^{{\rm s}},\\
\mathcal{F}^{\rm s} &\equiv& \frac{1}{2}\int \mu d{\mu}\,{f}^{{\rm s}},\\
\mathcal{S} &\equiv& -\left( \hat{j}+\hat{\chi}\right)\mathcal{E}^{\rm s}+\Sigma,
\end{eqnarray}
where $\mathcal{E}^{\rm s}$ and $\mathcal{F}^{\rm s}$ corresponds to the 
 radiation energy and flux of the streaming particle, and $\mathcal
S$ represents the source term that is a functional 
of the neutrino emissivity ($j$), absorptivity ($\chi$), and the 
isotropic diffusion term ($\Sigma$) all defined in the 
laboratory frame, respectively. From local hydrodynamic quantities
(density, $Y_e$, entropy), the source term of Eq.(1)
($\mathcal{S}[{j}, {\chi}, {\Sigma}]$) can be determined.
For closure, we use a prescribed relation between the radiation energy and
 flux as ($\mathcal{F}^{\rm s}/\mathcal{E}^{\rm s}= \frac{1}{2}(1 +\sqrt{1-
[{R_{\nu}}/\max\left(r,R_\nu\right)]^2})$ with $R_\nu$ being
 the radius of an energy-dependent scattering sphere 
(see Eq.(11) in \citet{idsa}). 
Since the cell-centered value of the flux, $\mathcal{F}^{\rm s}$, is
obtained by the prescribed relation,  the cell-interface 
value is estimated by the first-order upwind scheme assuming that 
the flux is out-going along the radial direction. 
With the numerical flux, the transport equation 
 of $\mathcal{E}^{\rm s}$ (Eq.(1)) now expressed in a hyperbolic form 
 is numerically solved.
 This modification does not produce any
 significant changes in the numerical results (see Takiwaki et al.
  in preparation for more details), however, 
the computational cost becomes more than 10 times smaller than that 
in the previous treatment. The velocity dependent terms ($O(v/c)$) are only
 included (up to the leading order) in the trapped part of the
 distribution function (Eq.
  (15) in \citet{idsa}).
 Concerning heavy-lepton neutrinos ($\nu_x = 
\nu_{\mu},\nu_{\tau},\bar{\nu}_{\mu}, \bar{\nu}_{\tau}$),
 we employ a leakage scheme to include the $\nu_x$ cooling
 via pair-, photo, plasma processes (e.g., \citet{ross03,itoh89}).
We apply 
the so-called ray-by-ray approach in which the neutrino transport is 
solved along a given radial direction assuming that the hydrodynamic 
 medium for the direction is spherically symmetric. 
To improve the accuracy of total energy conservation,
we follow the prescription proposed by \citet{mueller}.
For the calculations presented here, self-gravity is computed by 
 a Newtonian monopole approximation.
We use the equation of state (EOS) by \citet{latt91} with a compressibility modulus of $K = 180$ MeV (LS180).

 Our fiducial 3D models are computed on a 
spherical polar grid with a resolution of $n_r \times n_{\theta} \times
 n_{\phi}$ = $320 \times 64 \times 128 $, in which non-equally
 spacial radial zones covers from the center to an outer boundary of
 5000 km. The radial grid is chosen such that the resolution $\Delta
 r$ is better than 2km in the PNS interior and  typically better than
  5km outside the PNS. For the spectral transport, 
we use 20 logarithmically spaced energy bins ranging from 3 to 300 MeV
and we take a ray-by-ray approximation (e.g.,
\citet{buras06,bruenn13}), in which a ray is cast
for every angular zone. 
 In all the multi-D runs, 
the innermost 5 km is computed in spherical symmetry to avoid excessive
 time-step limitations. Seed perturbations for aspherical 
instabilities are imposed by hand at 10 ms after bounce by introducing
 random perturbations of $1\%$ in velocity behind the stalled shock.

 To test the sensitivity of the supernova dynamics to 
 numerical resolutions, we compute 3D model-series with lower
 angular resolutions, namely half or quarter of the (equidistant) 
mesh numbers in the 
 azimuthal direction ($n_r \times n_{\theta} \times
 n_{\phi}$ = $320 \times 64 \times 64$ and $n_r \times n_{\theta} \times n_{\phi}$ = $320 \times 64 \times 32$). In 2D simulations
 we vary the mesh numbers in the lateral direction as 
$n_r \times n_{\theta} $ = $320 \times 64$, $n_r \times n_{\theta} $ = $320 \times 128$ and  $n_r \times n_{\theta} $ = $320 \times 256$, respectively 
(see Table 1). In the table, model 3D-H-1 differs from model 
3D-H-2 (and 3D-H-3 etc) only in the random seed perturbations (with the 
 perturbation amplitudes being the same in all cases).
Note that the lowest-resolution 3D model in this work 
corresponds to the best-resolution model in \citet{takiwaki12}.
By using the fastest {\it K} computer in Japan, it took typically 
1.3 months (equivalently $\sim$ 4 million core-hour 
computing-time) for each of our 3D fiducial models.

\begin{deluxetable*}{lccccccccccccc}
\tablecaption{Model Summary}
\tablecolumns{12}
\tablewidth{0pc}
\tabletypesize{\scriptsize}
\tablehead{
\colhead{Model} & 
\colhead{$n_r \times n_{\theta} \times n_{\phi}$}    & \colhead{Angular} & 
\colhead{$t_{400}$}  & \colhead{ $ t_{400,{\rm av}}$} & \colhead{$\sigma[t_{400}]$} &
\colhead{$t_{\rm end}$} & \colhead{$R_{\rm max}$}    & \colhead{$E_{\rm diag}$}    &
\colhead{$(l,m)_{\mathrm{max}}$}  & \colhead{$|c_{l,m}|_{\mathrm{max}}$}   \\
\colhead{}      & 
\colhead{} & \colhead{resolution [$^{\circ}$]} &
\colhead{[ms]} & \colhead{[ms]}  &\colhead{[ms]}  &
\colhead{[ms]} & \colhead{[km]} & \colhead{[$10^{50}$ erg]} &
\colhead{} & \colhead{}
}
\startdata
3D-H-1  & $ 320 \times 64 \times 128 $ & $ 2.8^{\circ} \times  2.8^{\circ} $ & 223 & \multirow{3}{*}{236} & \multirow{3}{*}{24} & 284 &  550 & 0.15 & 1,0 & 0.02  \\
3D-H-2  & $ 320 \times 64 \times 128 $ & $ 2.8^{\circ} \times  2.8^{\circ} $ & 216 &                      &                     & 369 & 850  & 0.25 & 2,0 & 0.02  \\
3D-H-3  & $ 320 \times 64 \times 128 $ & $ 2.8^{\circ} \times  2.8^{\circ} $ & 269 &                      &                     & 269 &  400 & 0.15 & 2,0 & 0.02  \\
\noalign{\vskip 2pt}\hline\noalign{\vskip 2pt}
3D-M-1  & $ 320 \times 64 \times  64 $ & $ 2.8^{\circ} \times  5.6^{\circ} $ & 192 & \multirow{3}{*}{194} & \multirow{3}{*}{1.2}& 269 &  600 & 0.25 & 1,0 & 0.01  \\
3D-M-2  & $ 320 \times 64 \times  64 $ & $ 2.8^{\circ} \times  5.6^{\circ} $ & 194 &                      &                     & 319 &  700 & 0.30 & 1,0 & 0.02 \\
3D-M-3  & $ 320 \times 64 \times  64 $ & $ 2.8^{\circ} \times  5.6^{\circ} $ & 195 &                      &                     & 279 &  700 & 0.27 & 1,0 & 0.01  \\
\noalign{\vskip 2pt}\hline\noalign{\vskip 2pt}
3D-L-1  & $ 320 \times 64 \times  32 $ & $ 2.8^{\circ} \times 11.3^{\circ} $ & 193 & \multirow{3}{*}{188} & \multirow{3}{*}{4  }& 314 & 1000 & 0.60 & 2,1 & 0.009 \\
3D-L-2  & $ 320 \times 64 \times  32 $ & $ 2.8^{\circ} \times 11.3^{\circ} $ & 188 &                      &                     & 273 &  800 & 0.45 & 2,1 & 0.009  \\
3D-L-3  & $ 320 \times 64 \times  32 $ & $ 2.8^{\circ} \times 11.3^{\circ} $ & 183 &                      &                     & 273 &  700 & 0.35 & 2,1 & 0.01  \\
\noalign{\vskip 2pt}\hline\hline\noalign{\vskip 2pt}
2D-H-1  & $ 320 \times 256 $ & $ 0.7^{\circ} $ & 138 & \multirow{5}{*}{151} & \multirow{5}{*}{7} & 300 & 1100 & 0.65 & 1,0 & 0.05 \\
2D-H-2  & $ 320 \times 256 $ & $ 0.7^{\circ} $ & 154 &                      &                    & 329 & 1200 & 0.45 & 2,0 & 0.08 \\
2D-H-3  & $ 320 \times 256 $ & $ 0.7^{\circ} $ & 159 &                      &                    & 311 & 1200 & 0.76 & 2,0 & 0.1  \\
2D-H-4  & $ 320 \times 256 $ & $ 0.7^{\circ} $ & 156 &                      &                    & 368 & 1500 & 0.6  & 2,0 & 0.05 \\
2D-H-5  & $ 320 \times 256 $ & $ 0.7^{\circ} $ & 150 &                      &                    & 345 & 1200 & 0.7  & 2,0 & 0.1  \\
\noalign{\vskip 2pt}\hline\noalign{\vskip 2pt}
2D-M-1  & $ 320 \times 128 $ & $ 1.4^{\circ} $ & 140 & \multirow{5}{*}{142} & \multirow{5}{*}{10}& 369 & 1700 & 0.5  & 1,0 & 0.1  \\
2D-M-2  & $ 320 \times 128 $ & $ 1.4^{\circ} $ & 125 &                      &                    & 319 & 1800 & 0.8  & 2,0 & 0.06 \\
2D-M-3  & $ 320 \times 128 $ & $ 1.4^{\circ} $ & 151 &                      &                    & 400 & 1700 & 0.85 & 1,0 & 0.1  \\
2D-M-4  & $ 320 \times 128 $ & $ 1.4^{\circ} $ & 144 &                      &                    & 469 & 2300 & 1.0  & 1,0 & 0.1  \\
2D-M-5  & $ 320 \times 128 $ & $ 1.4^{\circ} $ & 152 &                      &                    & 400 & 1800 & 1.0  & 2,0 & 0.05 \\
\noalign{\vskip 2pt}\hline\noalign{\vskip 2pt}
2D-L-1  & $ 320 \times  64 $ & $ 2.8^{\circ} $ & 137 & \multirow{5}{*}{133} & \multirow{5}{*}{6} & 387 & 2100 & 1.1  & 2,0 & 0.08 \\
2D-L-2  & $ 320 \times  64 $ & $ 2.8^{\circ} $ & 137 &                      &                    & 395 & 2200 & 1.26 & 2,0 & 0.1  \\
2D-L-3  & $ 320 \times  64 $ & $ 2.8^{\circ} $ & 126 &                      &                    & 483 & 2800 & 1.3  & 1,0 & 0.09 \\
2D-L-4  & $ 320 \times  64 $ & $ 2.8^{\circ} $ & 140 &                      &                    & 559 & 2400 & 1.3  & 1,0 & 0.09 \\
2D-L-5  & $ 320 \times  64 $ & $ 2.8^{\circ} $ & 125 &                      &                    & 569 & 2500 & 1.3  & 2,0 & 0.05 \\
\noalign{\vskip 2pt}\hline\hline\noalign{\vskip 2pt}
\enddata
\label{tab:models}
\tablenotetext{}{Note that ``H'', ``M'', or ``L'' appended to our 
 models stands for high, moderate, low azimuthal angular resolutions, 
 respectively. The third number (-$i$) of each model that runs 
from 1 to 3 for 3D models and from 1 to 5 for 2D models
represents the difference only in the random seed perturbations 
(with the perturbation amplitudes being the same in all cases).
 $t_{400}$ represents the time when the average shock radius touches
  a radius of 400 km, and $t_{400,{\rm av}}$ and $\sigma[t_{400}]$
 denotes the model average
 and the dispersion of $t_{400}$. $t_{\mathrm{end}}$ denotes the time of the end of the simulation.
 $R_{\mathrm{max}}$ and $E_{\mathrm{diag}}$ are the averaged shock radius and 
the diagnostic explosion energy at $t_{\rm end}$.
 $(\ell,m)_{\rm max}$ denotes the set of spherical harmonics mode
  when the normalized amplitude takes the maximum
  value ($|c_{\ell,m}|_{\rm max}$) during the linear SASI phase.
}
\end{deluxetable*}

\section{Result}
As summarized in Table 1, all the computed models including nine 3D
models and fifteen 2D models exhibit shock revival, leading to
 the possibility of explosion.
 Before going into detail how the explosion
dynamics and stochasticity are different in 2D and 3D, we briefly
outline the hydrodynamics features taking model 3D-H-1 as an example.
   
The top panel of Figure \ref{fig1} shows the blast morphology of
 model 3D-H-1 at $t_{\rm pb} = 230$ ms (postbounce) when the
  revived shock is reaching an angle-averaged radius of 400 km
  (e.g., red dashed line in the right panel of Figure. \ref{fig2}). 
As seen from the side wall panels, a bipolar
 explosion is obtained for this model. The bottom left panel
  (red regions) shows that the ratio of the residency timescale to
  the neutrino heating timescale (e.g., Equations (6) and (7) in 
   \citet{takiwaki12}) exceeds unity behind the shock, which presents
    evidence that the shock revival is driven by the neutrino-heating
     mechanism. The bottom right panel of Figure \ref{fig1} depicts 
spacial distribution of the net neutrino heating rate at
$t_{\rm pb} = 150$ ms. Small scale inhomogeneities (colored as red or
yellow) are seen, which predominantly comes from
 neutrino-driven convection and anisotropies of the accretion flow,
 but the shape of the gain region is very close to be spherical
before the onset of an explosion.
This suggests that the bipolar geometry of the shock is
produced not by the global anisotropy of the neutrino heating
in the vicinity of the neutrino sphere, but by multi-D
 effects such as by neutrino-driven convection and the SASI in the gain
  regions after the explosion (gradually) sets in.

  Reflecting the stochastic nature of the multi-D neutrino-driven
  explosions, the blast
    morphology changes from models to models.
The left and middle panels of Figure \ref{fig2} show that a stronger
 explosion is obtained toward the north direction (model 3D-H-2)
  and the south pole (model 3D-H-3), which is only different
   from model 3D-H-1 (e.g., Table 1) in terms of the imposed
   initial random perturbations. Note that due to the use of the spherical
    coordinates, we cannot omit the possibility
    that the polar axis still gives
     a special direction in our 3D simulations. But more importantly,
      our results show that the final fate of the 3D and 2D models
      whether an explosion is obtained or not, is little affected
      by the stochasticity of the explosion geometry.

       In fact, the right panel of Figure \ref{fig2} shows the evolution of the average
shock radius for our 1D (blue line), 2D (green lines), and 3D (red lines)
 models, respectively. Before the onset of shock revival (before 100 ms after bounce),
 the evolution of the shock is all similar to that of the 1D model (blue
 line). After that, our results show that
 the shock expansion is systematically more energetic in 2D (green
 lines) than in 3D (red lines).
 This feature is qualitatively consistent with \citet{takiwaki12}, and also with \citet{hanke13} who recently reported
  2D vs. 3D comparison
    based on a single 3D model but employing more 
detailed neutrino transport than ours.

\begin{figure}[htb]
\centering
\includegraphics[width=.8\linewidth]{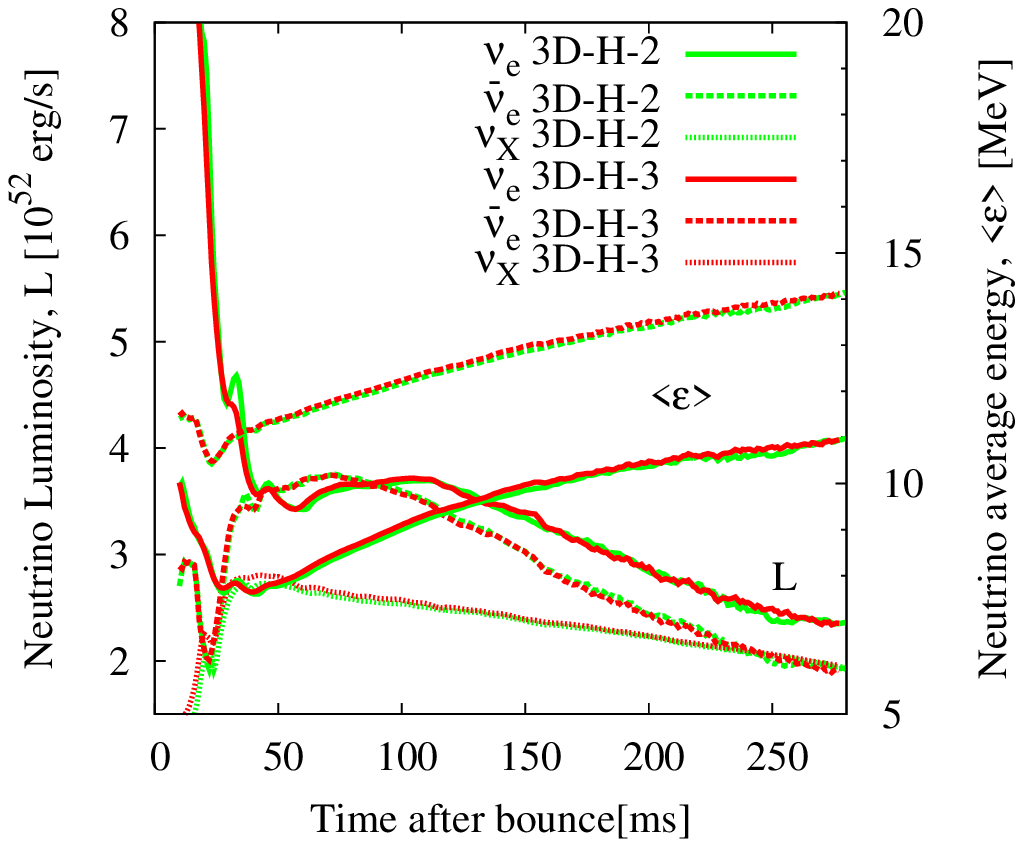}
\includegraphics[width=.8\linewidth]{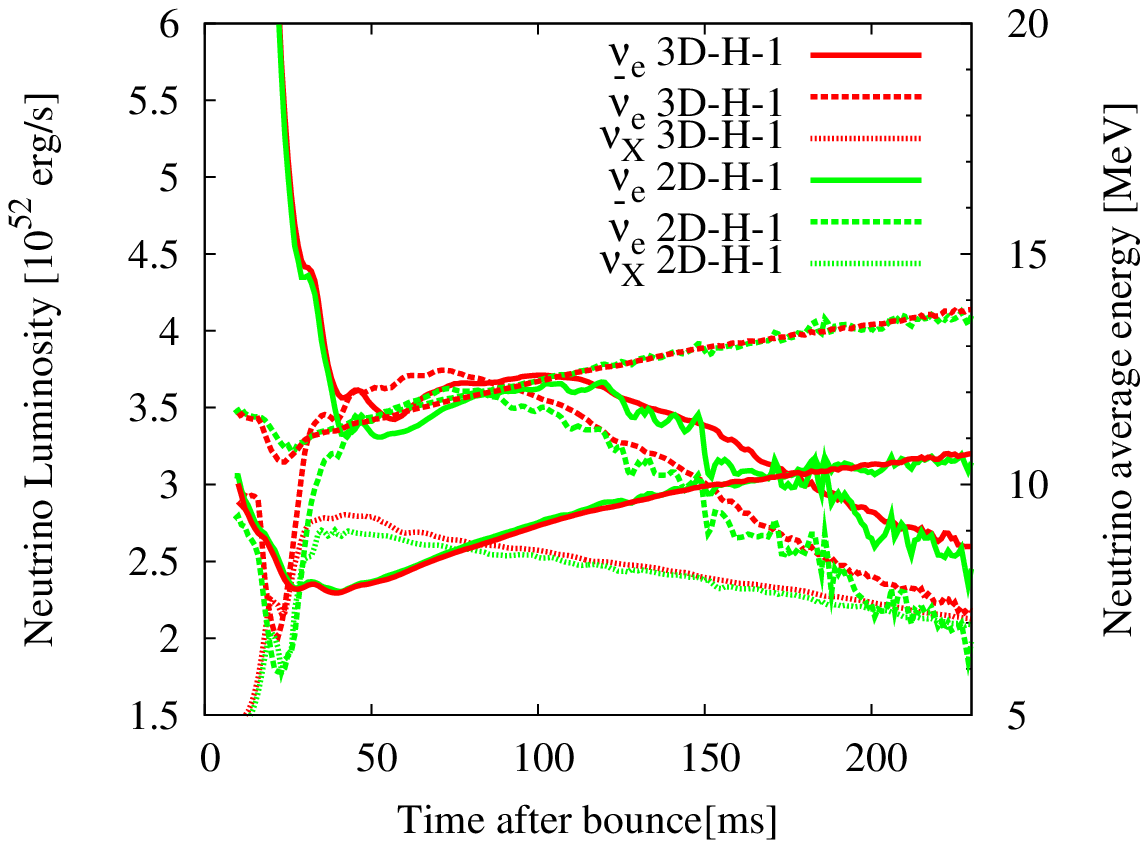}
\caption{Top panel shows time evolution of neutrino luminosities and mean energies
  of electron ($\nu_e$), anti-electron ($\bar{\nu}_e$), or
 heavy-lepton ($\nu_X$) neutrinos for models 3D-H-2 (green line) and 3D-H-3 (red line), respectively. Bottom panel is the same as the top panel 
 but for the comparison between 2D and 3D (for models 3D-H-1 
and 2D-H-1). These quantities are estimated at 500 km.}\label{fig3}
\end{figure}

Due to our lack
   of necessary computational resources,
    our 3D models should be terminated typically before $t_{\rm end} \lesssim 300$ ms
 postbounce (e.g., in Table 1), but we expect
     them to produce explosions subsequently, seeing
      a continuous shock expansion out to
     a radius of 500 km in 3D (and 700 km in 2D).
       Given the same numerical resolution
       (e.g., model series 3D-L in Table 1), the average shock radii
 in this study is smaller than those in \citet{takiwaki12}, in which the cooling
   by heavy-lepton neutrinos was not taken into account.
Due to the inclusion of the $\nu_x$ cooling, the (angle-averaged) 
$\bar{\nu}_e$ luminosity decreases more quickly after bounce 
(compare Figure
   \ref{fig3} and Figure 14 of \cite{takiwaki12}), which leads
    to the less energetic shock expansion in this study.
   It should be mentioned that by comparing
     our $\nu_x$ luminosity estimated by the leakage scheme with that
     obtained by the work by \citet{buras06} with detailed neutrino transport,
     the peak luminosity is more than 20 \% smaller in our case. Such
     underestimation of cooling by heavy-lepton neutrinos should lead to artificially
     larger maximum shock extent ($R_{\rm max} \sim 260$ km,
     blue line in the right panel
      of Figure \ref{fig2}) compared to $R_{\rm max} \sim170$ km in
 \citet{buras06}. We have to emphasize that the use of the leakage scheme
 together with
 the omission of inelastic neutrino scattering on electrons and
 GR effects in the present scheme is likely to facilitate
 artificially easier explosions. Regarding our 2D models, the relatively
 earlier shock revival ($\sim 100$ ms postbounce) coincides with 
 the decline of the mass accretion rate onto the central PNS.
 This could be the reason that the timescale is similar to that 
in \citet{bernhard12} who reported 2D (GR) models for
 the same progenitor model with detailed neutrino transport.

  As seen from Figure \ref{fig3}, the angle average neutrino luminosity ($\langle L_{\nu} \rangle$)
 and the mean neutrino
  energy ($\langle \epsilon_{\nu} \rangle = \int E^3 \mathcal{F}^{\rm s}dE  /\int E^2 \mathcal{F}^{\rm s}dE$, where $E$ is neutrino energy)
  is barely affected by the imposed initial perturbations (presumably at
    few-percent levels in amplitudes).
 This again supports
   our finding that the explosion stochasticity is very
   influential to determine the blast morphology but not the working of
    the neutrino-heating mechanism. 

From the bottom panel of Figure \ref{fig3}, it can be seen
 that overall trend in the neutrino luminosities and 
the mean energies is similar between our 3D and 2D model.
 The neutrino luminosities in the 2D model (green lines) 
show a short-time variability
 (with periods of milliseconds to $\gtrsim 10$ ms) after around 100 ms 
postbounce. Such fast variations in the postbounce luminosity 
evolution have been already found in previous 2D studies (e.g., 
\citet{ott_multi,marek_gw}). This is caused by the modulation 
of the mass accretion rate due to convective plumes and downflows
 hitting onto the PNS surface (see also \citet{lund12, tamborra13} 
about the detectability of these neutrino signals). It is interesting to note that 
such fast variability is less pronounced in our 3D
 model (red lines in the bottom panel). This is qualitatively consistent 
with \citet{lund12} who analyzed the neutrino signals from 2D and 3D 
models, in which an approximate neutrino transport was solved 
\citep{annop10} as in \citet{scheck06}.

\begin{figure}[htb]
\centering
\includegraphics[width=.8\linewidth]{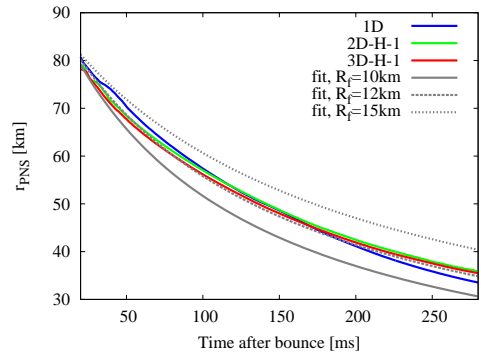}
\caption{Average PNS radius (defined by a fiducial density of  
 $10^{11}~ {\rm g}~{\rm cm}^{-3}$) for the 1D (blue line), 
2D (green line), and 3D models (red line), respectively. 
The gray lines labeled as "fit" 
corresponds to the prescribed PNS evolution with three different values 
of $R_{\rm f} = 10, 12, $ and 15 km (see text for more details).
}\label{pns}
\end{figure}

Figure \ref{pns} shows evolution of the average PNS radius
 for the 1D (blue line), 2D (green line), and 3D models (red line),
 respectively, that are defined by a fiducial density of  
 $10^{11}~ {\rm g}~{\rm cm}^{-3}$.
  The PNS contraction is similar among the 1D, 2D, and 3D models.
  Although the PNS contraction potentially affect the evolution of the shock \citep{hanke13,suwa13}, in our cases that are unchanged by the difference of the dimension and that are not main agent to explain the divergence of the shock evolution in 1D, 2D and 3D. 
  The PNS contraction is slightly stronger
 in the later postbounce phase in 1D ($\gtrsim 150$ ms postbounce, 
compare blue with green and red lines)) compared to 2D and 3D 
 because  no shock revival was obtained in the 1D model and heavier PNS and slightly deeper 
 gravitational potential are obtained compared to that of the multi-D models.
 In the figure,
 three more lines (solid, dashed, dotted gray lines) are plotted, 
in which we estimate the evolution of the PNS radius based on
 the fitting formula (equation (1) of \citet{scheck06}) 
by changing a final radius of PNS $R_{\rm f}$ for a given set of an 
exponential timescale of $t_{\rm ib} = 1$ s and an initial radius of PNS 
$R_{\rm i} =$ 85 km. As can be seen, the dashed
 gray line ($R_{\rm f} = 12$ km) can most closely reproduce our results,
 which is just between the slow and fast contraction investigated in the 
 work by \citet{hanke13}.

\begin{figure}[htb]
\centering
\includegraphics[width=.8\linewidth]{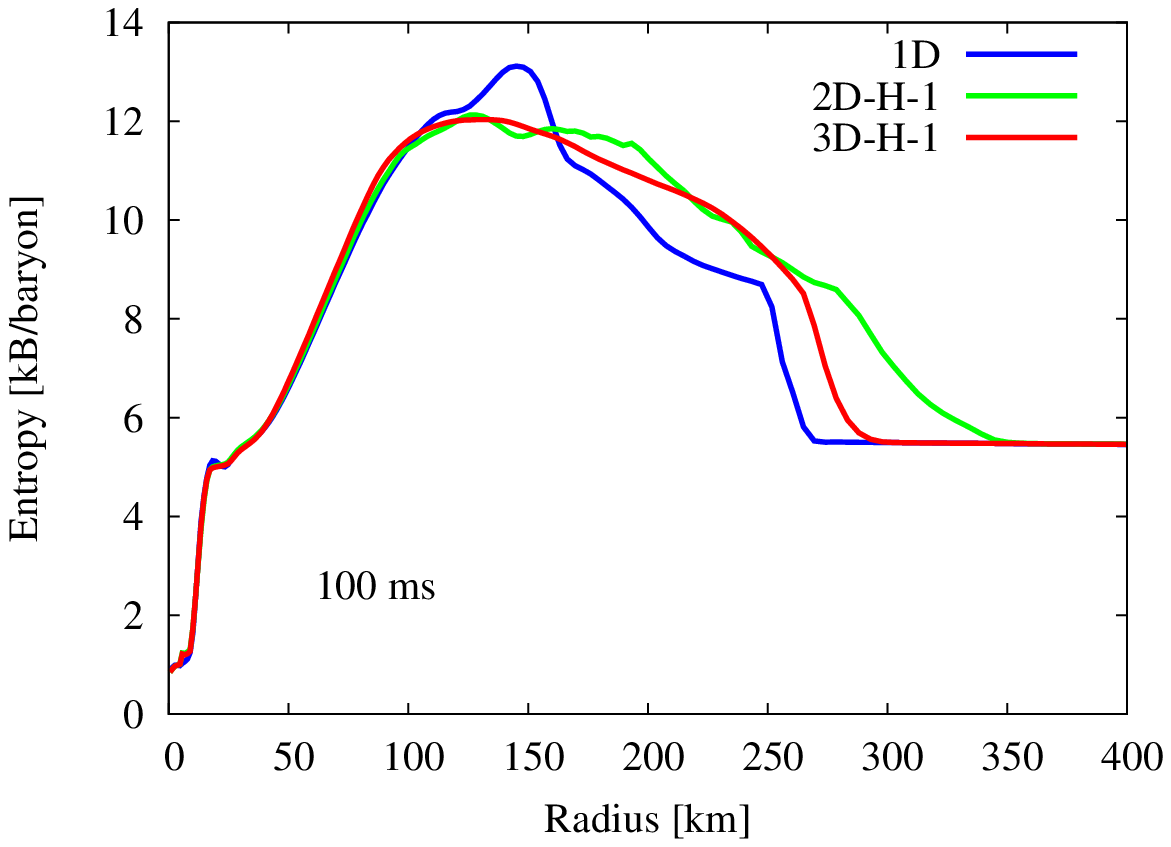}\\
\includegraphics[width=.8\linewidth]{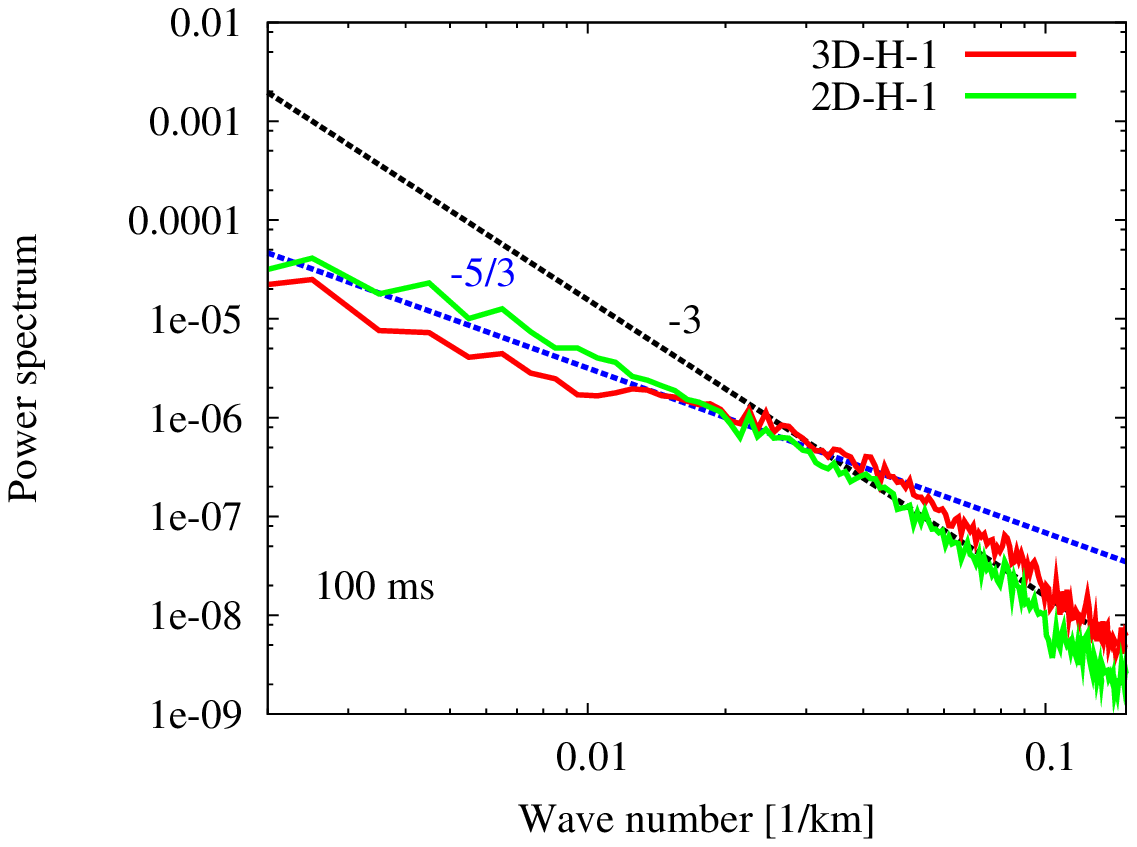}\\
\caption{Top panel shows radial profiles of angle-averaged entropy at 
100 ms postbounce for the 1D, 2D-H-1, 3D-H-1 model, respectively.
 Bottom panel shows spectra of the 
  turbulent kinetic energy as a function of wavenumber ($k$) 
between our 2D (green line) and 3D (red line) model, respectively.
}\label{fig:r-s} 
\end{figure}

Top panel of Figure \ref{fig:r-s} shows angle-averaged entropy profile at 100 ms postbounce,
 after when the difference of the subsequent shock evolution between our 
1D, 2D, and 3D models becomes remarkable (e.g., right panel in Figure \ref{fig2}).
As has been studied in detail since 1990's (e.g., \citet{herant94,burrows95,jankamueller96}), buoyancy-driven convection 
supported by turbulence (e.g., \citet{murphy13}) transports heat 
radially outward, leading to a more extended entropy profile
 in the 2D (green line in the panel) and 3D (red line) model 
compared to the 1D model (blue line) (see also \citet{hanke12}).

Bottom panel of Figure 5 compares the turbulent energy spectra of the 
anisotropic velocity (\cite{takiwaki12}) 
 as a function of wavenumber ($k$) between our 2D (green line) and 
3D (red line) model, respectively. The green and red line crosses at 
around $k_{\rm cross}$ = 0.02/km (corresponding to $\sim$ 50 km 
in spacial scale), above which the amplitude for the 3D model (red line) 
dominates over that for the 2D model (green), below which the amplitude 
for the 2D model does for the 3D model. This is qualitatively 
consistent with the previous results using the light-bulb method (e.g., 
\citet{hanke12,burrows12,dolence12,couch13,fernandez13}. 
By comparing the 2D and 3D curve with spectral slopes labeled by the 
corresponding exponent (-5/3:dotted 
blue line, -3:dotted black line), the power-low dependence 
($\propto k^{-5/3}$) approximately holds for $k_{\rm int} \lesssim 0.02$ 
/km in the 2D model (green line) presumably as a result of the inverse 
energy cascade \citep{kraichnan}. Above $k_{\rm cross}$, the slope of the 
3D model (red line) become more closer to $k^{-5/3}$ (blue line), 
while the spectrum slope of the 2D model drops much steeply with
 the wave number as $k^{-3}$ (gray line). These features are again in 
accord with the previous studies mentioned above
(e.g., \citet{hanke12,burrows12,dolence12,couch13,fernandez13}).

As shown in the middle panel of Figure \ref{fig2},
  model 3D-H-3 produces one-sided explosion
  towards the south pole during the simulation time.
Reflecting the unipolarity,
the average shock radius is smaller than for the other 3D models (compare
 dotted line (in red) with solid and dashed red (in red) in the right
 panel of Figure \ref{fig2}).
To quantify the vigor of the shock expansion ,
 $t_{\mathrm{400}}$ is a useful quantity that was defined
 in \cite{hanke12} as the moment of time when the shock
  reaches an average radius of 400 km. In fact, as seen
   from Table 1, $t_{\mathrm{400}}$ of model 3D-H-3
   ($t_{\mathrm{400}}
     \approx 270~\mathrm{ms}$) is delayed about 50 ms compared to
     those of models 3D-H-1 and 3D-H-2 ($t_{\mathrm{400}}
     \approx 220~\mathrm{ms}$).

     The model average of $t_{400}$ (e.g., $t_{400,{\rm av}}$ in
     Table 1) clearly shows that models with higher numerical
     resolutions lead to slower onset of the shock revival
      in both our 3D and 2D models.
      This feature is qualitatively consistent with the
      2D self-consistent models by \cite{marek} and
      with the 3D idealized models by \cite{hanke13, couch13},
      respectively.

$\sigma_{400}$ in Table 1 represents the model dispersion of $t_{400}$, which
 varies much more stochastically in 3D models with different numerical
 resolution (from 1.2 to 24 ms) than those in 2D (from 6 to 10 ms).
 For model 3D-H-3,  the shock revival is most delayed (e.g., Table 1)
 and the shock expansion is weakest among the computed models (dotted
  red line in Figure \ref{fig1}).
\citet{nagakura13} proposed that shock revival 
 is very sensitive to the imposed 
seed perturbations near the stalled shock. Following the hypothesis, we speculate that
 the influence of seed perturbations is seen most remarkably
 in our weakest explosion model.

 $E_{\rm diag}$ in Table 1 denotes the diagnostic energy
 defined as the total energy (internal plus kinetic plus
  gravitational), integrated over all matter where the sum of the corresponding
    specific energies is positive. We include recombination energy in internal energy(\cite{bruenn13}).
 Reflecting the earlier shock revival,
 the diagnostic energy is systematically bigger in 2D than in 3D.
 These diagnostic energies when we terminated the simulation were
   typically on the order of $\sim 10^{49}$ erg and $\sim 10^{50}$ erg for our
   3D and 2D models, respectively. 
It should be noted that this quantity is estimated at the end of simulation,$t_{\rm end}$. In order to compare
   the {\it diagnostic} energy with the observed kinetic explosion
 energy ($\sim 10^{51}$ erg), a much longer-term
 simulation including improved microphysics,
general relativity, and
 nuclear burning would be needed.

\begin{figure*}[htb]
\centering
\includegraphics[width=.4\linewidth]{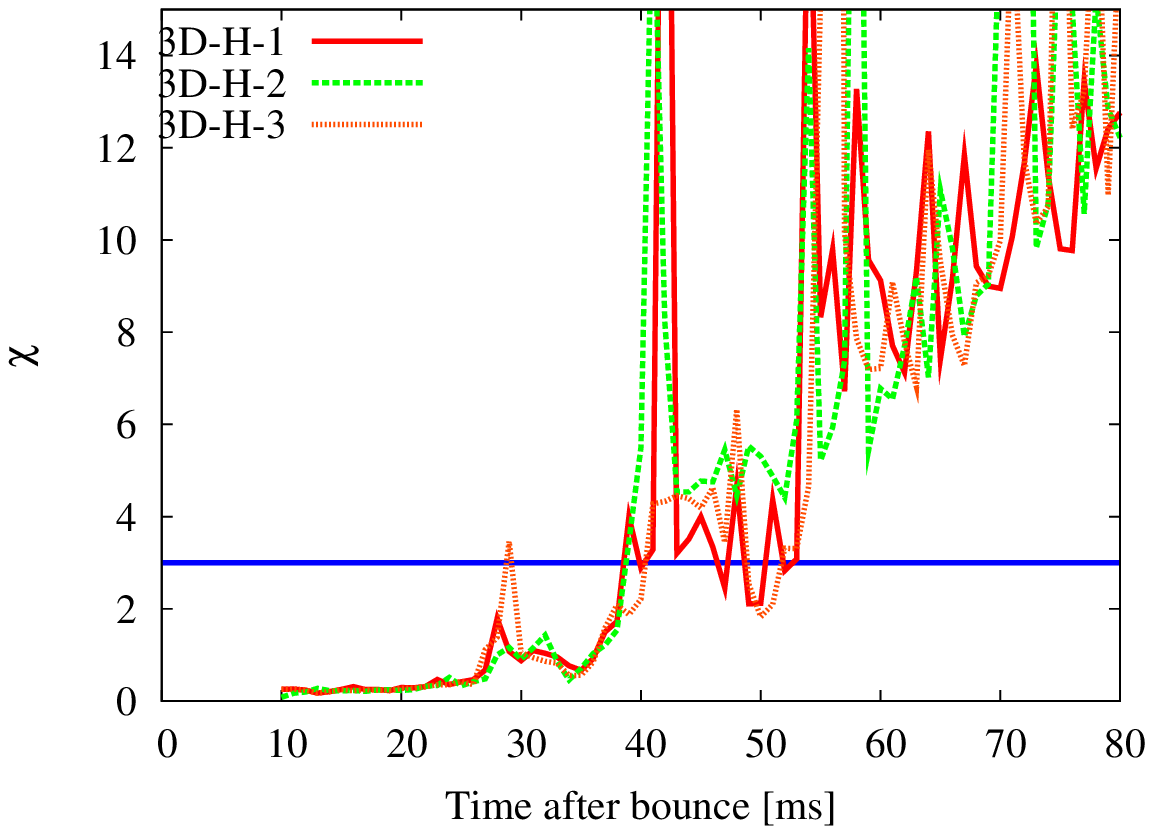}
\includegraphics[width=.4\linewidth]{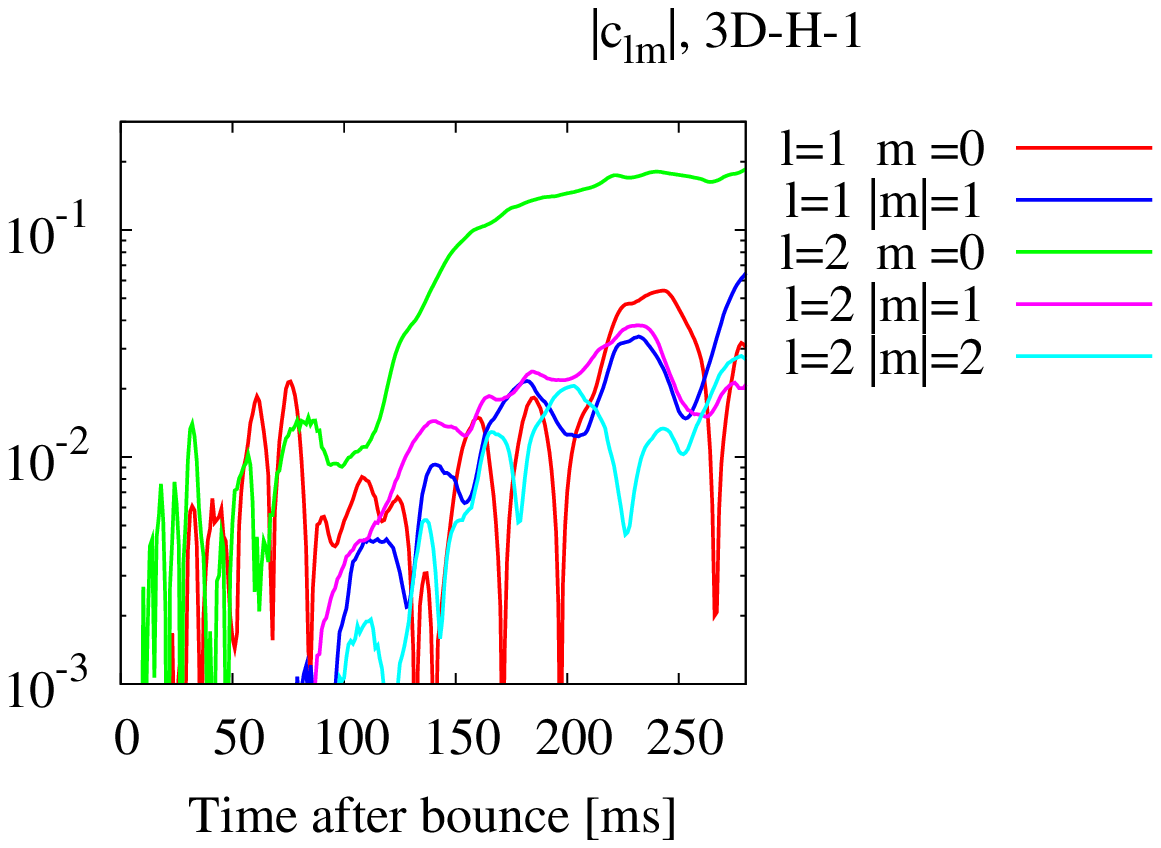}
\caption{Time evolution of $\chi$ parameter (left panel) and
  normalized spherical harmonic mode amplitudes of the shock for our 3D
  high-resolution models with respect to each $(\ell, m)$ (right panel,
  e.g., $|c_{\ell m}/c_{00}|$ in equation (10) of \citet{takiwaki12}).
 The horizontal blue line corresponds to $\chi=3$, above which
 neutrino-driven convection dominates over the SASI \citep{thierry}. 
In the right panel, lower modes with larger amplitudes
 are selected.}
 \label{fig5}
\end{figure*}

 Recently it is enthusiastically discussed which
one between neutrino-driven convection and the SASI plays
 a more crucial role in facilitating neutrino-driven explosions \citep{thierry,dolence12,murphy13,burrows12,hanke12,hanke13}.
  The left panel of Figure \ref{fig5} shows the time evolution of the
Foglizzo parameter $\chi$ \citep{thierry}
 As seen, $\chi$ continuously exceeds
 the critical number of $3$ \citep{thierry} rather shortly after 
 bounce ($\sim 40$ ms), marking the transition to the 
 non-linear phase. The earlier onset of the shock revival 
 and the absence of clear features of the SASI in the 
 linear phase (see discussion below) could indicate that 
 neutrino-driven convection dominates over the SASI when
 the explosion sets in (e.g., at $t_{400}$ in Table 1).

The right panel of Figure \ref{fig5} shows the evolution of
 normalized coefficient of spherical harmonics of the shock surface
 for model 3D-H-1. No clear feature of the 
 linear growth of the SASI ($\lesssim 40$ ms postbounce) 
was obtained for the 11.2 $M_{\odot}$ star explored in this work. 
In both our 3D and 2D models, the qualitative behaviors of the harmonic
 modes seen in Figure 6 are less sensitive to the employed numerical 
resolution and 
seed random perturbations (see appendix A).
 The dominant channels are of low modes (e.g., $(\ell,
 m)_{\rm max}$ in Table 1), the amplitudes of which (e.g.,
 $|c_{\ell,m}|_{\rm max}$ in the table) are
 systematically larger in 2D than in 3D. This is qualitatively
 consistent with previous
   3D simulations employing different numerical setups (e.g.,
 \citet{nordhaus,hanke12,burrows12,hanke13}). 

After the linear phase
 comes to an end at around 100 ms postbounce, the trajectory of the
 revived shock shows a wider diversity (e.g., Figure \ref{fig1}) depending
  on the employed numerical resolution and seed perturbations.

\section{Summary and Discussion}
Studying an 11.2 $M_{\odot}$ and changing numerical resolutions and seed perturbations systematically
 in the multi-D simulations employing the updated IDSA scheme,
 we studied how the postbounce dynamics is different in 2D and 3D. 
  All the computed models exhibit the
 neutrino-driven revival of the stalled bounce shock, leading to the
 possibility of an explosion. Though the blast morphology changes from
 models to models reflecting the stochastic nature of multi-D
 neutrino-driven explosions, it was found that the final fate of these
 multi-D models whether an explosion is obtained or not, is little 
affected by the explosion stochasticity at least in the current investigated progenitor model. In line with some previous
 studies, higher numerical resolutions lead to slower onset of the 
shock revival in both 3D and 2D.  Our results systematically 
showed that the revived shock expands more 
energetically in 2D than in 3D.

      The caveats of our 3D models
       include the ray-by-ray approximation, the use of the softer
        EOS, and the omission of detailed neutrino reactions and general
       relativity (e.g., \citet{kuroda12,ott13}). Keeping our efforts to improve them,
       it is important to study the dependence of
        progenitors (e.g., \citet{buras06,bruenn13}) and EOS
        (e.g., \citet{marek,suwa13}) on the neutrino-driven mechanism in 3D
        computations. A number of exciting issues also remain to be
       investigated,
        such as gravitational-wave signatures (e.g.,
       \citet{kotake13}),
       neutrino emission and its
 detectability (e.g., \citet{lund12}), and the possibility of 3D SASI flows generating pulsar
  kicks and spins (e.g., \citet{annop13}). Shifting from individuals to
 populations of 3D models, a rush of 3D explorations with increasing
  sophistication is now going to shed light on these fascinating 
riddles (hopefully not in the distant future) with increasing 
supercomputing resources on our side.

{\acknowledgements 
We are thankful to K. Sato and S. Yamada for continuing encouragements.
Numerical computations were carried on in part on the K computer at the 
RIKEN Advanced Institute for Computational Science (Proposal number 
hp120285), XC30/XT4 at NAOJ, and on SR16000 at YITP in Kyoto University.
 This study was supported in part by Grants-in-Aid for Scientific
 Research (Nos. 23540323, 23340069, 24244036, and 25103511) and by HPCI Strategic
 Program of Japanese MEXT.}

\appendix

\section{Dependence on the amplitude of seed perturbations}

 In this Appendix we briefly report on the dependence on the 
amplitudes of the initial seed perturbations.
We have added a radial velocity perturbation, $\delta v_r(r,\theta,\phi)$ to the profile obtained by 1D simulation,$v_r^{1D}(r)$, according to the equation
$\delta v_r = p_{\rm amp} {\rm rand} \times v_r^{1D}$
where rand is pseudorandom number that takes the value from $-1$ to $1$ and
$p_{\rm amp}$ represents the 
 absolute amplitude of the seed perturbations that we take as $1 \%$ in the models discussed in the main section.
Let us firstly note that the amplitude of seed perturbations $a_{\ell,m}$ 
assumed in this study scales as $p_{\rm amp}/N^{1/2}$,
where $a_{\ell,m}$ denotes the amplitude of the $(\ell,m)$ component of 
the seed perturbations at a given radius (i.e. $a_{\ell,m} = \int d \phi \int \sin \theta d \theta \left(v_r^{1D}+\delta v_r \right)Y_{\ell,m}^{*}$, where $Y_{\ell,m}^{*}$ is conjugate of spherical harmonic function) 
 and $N = n_{\theta} \times n_{\phi}$ 
represents the total angular mesh number where
$n_{\theta}, n_{\phi}$ being the mesh number in the $\theta$ and $\phi$ direction, respectively(e.g. 
in our 3D model with highest resolution, $n_{\theta} = 64, n_{\phi} = 128$).
Therefore the seed 
amplitudes assumed in this work depend on numerical resolution 
(i.e., $N$).

\begin{figure*}[htb]
\centering
\epsscale{0.4}
\plotone{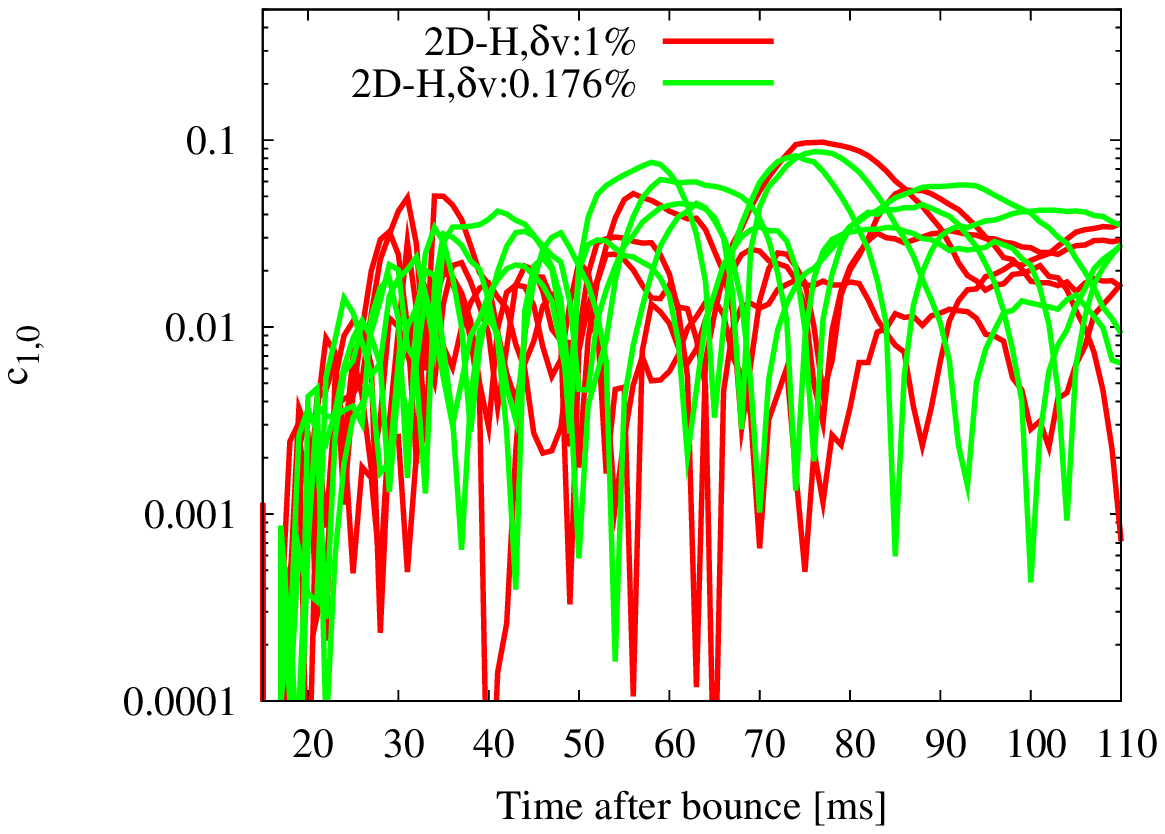}\\
\epsscale{0.8}
\plottwo{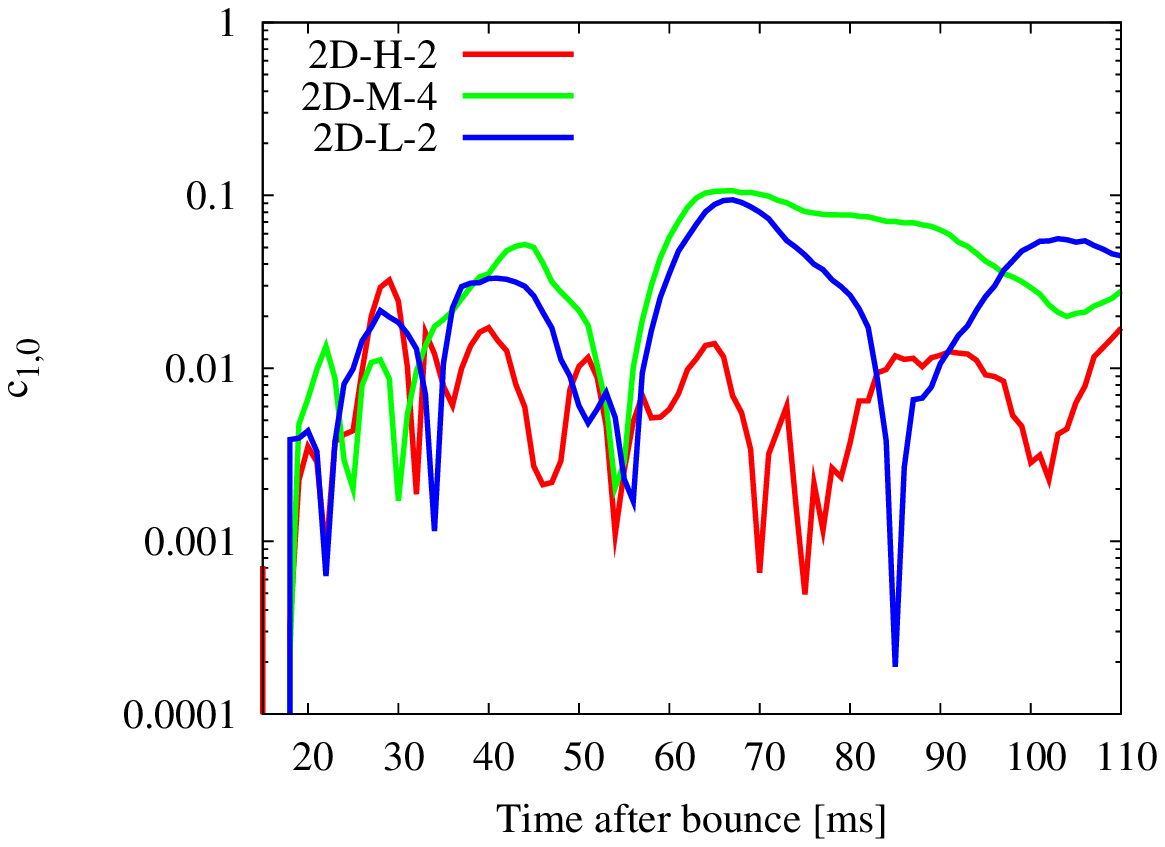}{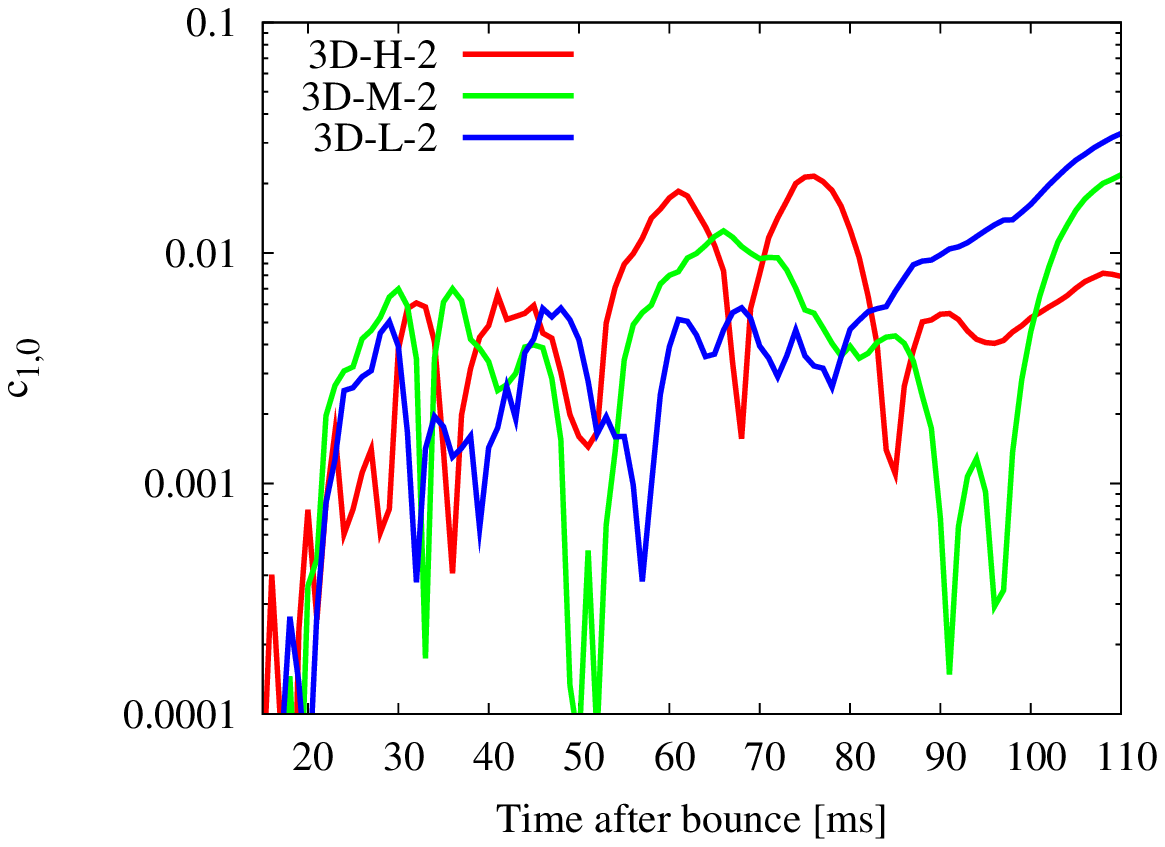}
 \caption{Top panel shows evolution of $c_{1,0}$ 
 for our 2D models with $p_{\rm amp} = 1\%$ (red lines) 
or $0.176 \% (= 1\%/\sqrt{32})$ (green lines) with 5 different 
realizations of random perturbation (see text for more details). 
Bottom panels show evolution of $c_{1,0}$ for different resolution (H, M, and L) for our 2D and 3D models.}
 \label{app:fig1}
\end{figure*}

If we keep the seed amplitudes ($a_{\ell,m}$) constant for models 
with different resolutions, will our results change 
drastically? Since we cannot rerun 3D models due to 
 the limited computational resources, we compute a number of 2D models
to answer to this question. First of all, let us discuss 2D models
 with different seed amplitudes. Top panel of Figure \ref{app:fig1} 
shows the evolution of $c_{1,0}$ (the normalized harmonic amplitudes 
of shock position) for 2D models with $p_{\rm amp} = 1\%$ (red lines) or
 $0.176 \% (= 1\%/\sqrt{32})$ (green lines) with five different realizations 
of initial random perturbation.
The reduced amplitude is determined by the ratio of total grid number of 2D-H models and 3D-H models (i.e.  $1/\sqrt{32} = \sqrt{
n_\theta |_{2D-H}/(n_\theta n_\phi)|_{3D-H}}$).
 One might have a guess that $c_{\ell,m}(t)$ could 
evolve as $a_{\ell,m} \exp(t/t_0)$ with $t_0$ representing the duration 
of the linear growth rate of hydrodynamic instabilities including 
the SASI or neutrino-driven convection. If this could be the case, 
the linear growth amplitude (at before 35 ms postbounce in the top panel) should be higher about $\sqrt{32}$ times for red lines compared 
to green lines. But as it is shown, this is not the case\footnote{
Note that if the initial seed perturbations would be more bigger than 
those assumed in this study, they should affect the postbounce 
hydrodynamics (see \citet{couch_ott13})}.
 And after the early rising phase (about 40 ms postbounce in the panel),
 the saturation amplitudes are shown to be insensitive to the initial 
perturbation amplitudes (for the initial strength employed above). 
Remembering that neutrino-driven convection is likely to dominate
 over the SASI in the non-linear regime for the 11.2 $M_{\odot}$ 
progenitor employed in this work,  the delay of $t_{400}$ (in
 the non-linear phase)
for models with higher numerical resolutions cannot be also simply 
ascribed to the difference of the seed amplitudes.

Bottom panels of Figure \ref{app:fig1} show evolution of $c_{1,0}$ for 
different resolution (H, M, and L) for
 our 2D and 3D models, respectively. For a given numerical resolution, 
$t_{400}$ of the chosen models in these plots is close to a median of 
$t_{400}$ for each model series. Note that in these models the initial 
seed amplitudes are dependent on both resolution and dimension. 
 As can be seen, 
in both 2D and 3D, the amplitudes in the linear phase ($<40$ms postbounce) are comparable for models with different resolutions.
 And in the non-linear regime, no monotonic dependence of the non-linear evolution  on the initial seed amplitudes (between H, M, L models)
can be found. These results show that the findings in this work
are less sensitive to the assumed 
initial seed perturbations.

\end{document}